\documentstyle[12pt]{article}

\begin{document}
%\begin{titlepage}

\begin{center}
%{\hbox to\hsize{
%\hfill \bf hep-ph/??? }}
{\hbox to\hsize{\hfill December 2007 }}

\bigskip \vspace{3\baselineskip}

{\Large \bf Noncommutative  corrections to classical black holes  \\ }

\vspace{2 cm}

{\bf Archil Kobakhidze  \\ }

\smallskip

{ \small \it School of Physics, Research Centre for High Energy Physics,\\ 
The University of Melbourne, Victoria 3010, Australia \\ 
 E-mail: archilk@unimelb.edu.au  \\}

\bigskip \vspace*{.5cm}

{\bf Abstract}\\

\end{center}

\noindent 

{\small 
We calculate leading long-distance  noncommutative corrections to  the classical 
Schwarzschild black hole which is sourced by a massive noncommutative scalar field. The energy-momentum tensor is taken up to ${\cal O}(\ell^4)$ in noncommutative parameter, and is treated in semiclassical (tree level) approximation.  These noncommutative corrections can dominate classical post-post-Newtonian corrections providing $\ell > 1/M_P$, however, they are still too small to be observable in present-day experiments.  }

\bigskip

\bigskip

%\end{titlepage}

\baselineskip=18pt
\paragraph{I.} 
Recently, a number of different attempts \cite{Chamseddine:2000si}-\cite{Kobakhidze:2006kb} have been made to formulate a theory of gravitation 
on canonical noncommutative space-time defined through the following algebra of coordinates: 
\begin{equation}
\left[\hat x^{\mu}, \hat x^{\nu}\right ]=i\ell^2\theta^{\mu\nu}~.
\label{1}
\end{equation}
Here $\ell$ is a length parameter of noncommutativity and $\theta^{\mu\nu}$ is a constant antisymmetric matrix.\footnote{Other approaches to noncommutative gravity can be found in recent reviews \cite{Szabo:2006wx}. See also references therein. }  The essential difference between these formulations of noncommutative gravity lies  in the  treatment of symmetries of general relativity in noncommutative setting.  The resulting actions contain rather complicated noncommutative corrections to the Einstein-Hilbert Lagrangian of classical general relativity, and thus it is rather difficult to analyze corresponding equations of motions.   It has been observed though [see also \cite{Mukherjee:2006nd}], that in all these formulations the lowest order noncommutative corrections are ${\cal O}(\ell^4)$.  Thus in the leading approximation pure gravity action can be treated undeformed.

Our aim in this paper is to calculate noncommutative long-distance corrections to classical black holes.\footnote{Noncommutative corrections ${\cal O}(\ell^4)$ to classical black hole metrics has been recently calculated in \cite{Chaichian:2007we} within a specific theory \cite{Chamseddine:2000si} of noncommutaive gravity, by solving truncated equations of motion.}  We follow the formalism of the effective field theory which has been applied to calculate commutative quantum corrections to classical black holes in \cite{Donoghue:2001qc}.  According to the discussion in the previous paragraph, the leading interactions of the linearized graviton field with a matter are given by the Lagrangian, 
\begin{equation}
{\cal L}_{\rm int}=\int d^4x\frac{1}{2}h_{\mu\nu}T^{\mu\nu}_{\rm NC}~,
\label{2}
\end{equation}
 where $T^{\mu\nu}_{\rm NC}$ is the noncommutative energy-momentum tensor which can be expanded in series of noncommutative parameter $\ell^2$, 
 \begin{equation}
 T^{\mu\nu}_{\rm NC}=T^{\mu\nu}_{0}+T^{\mu\nu}_{1}+T^{\mu\nu}_{2}...
 \label{3}
 \end{equation}
 n$^{\rm th}$ term in this expansion is of the order of ${\cal O}(\ell^{2n})$, that is, $T^{\mu\nu}_{0}$ is the usual commutative energy momentum tensor.   The graviton field for nearly static source can be solved as (we work in the harmonic gauge, $\partial^{\mu}(h_{\mu\nu}-\frac{1}{2}\eta_{\mu\nu}h)=0$):
 \begin{equation}
 h_{\mu\nu}(x)=-16\pi G_N \int \frac{d^3\vec q}{(2\pi)^3}e^{i\vec q\vec r}\frac{1}{\vec q^2}\left (T^{\mu\nu}_{\rm NC}(q)-\frac{1}{2}\eta_{\mu\nu}T_{\rm NC}(q)\right )~,
 \label{4}
 \end{equation}
 where $T_{\rm NC}=\eta_{\mu\nu}T^{\mu\nu}_{\rm NC}$, and $T^{\mu\nu}_{\rm NC}(q)=\langle p_2|:T^{\mu\nu}_{\rm NC}(x):|p_1\rangle$, $q_{\mu}=(p_2-p_1)_{\mu}$. Once the explicit form of energy-momentum tensor is known, the matrix elements can be calculated perturbatively. In what follows we restrict ourselves  to the lowest semiclassical (tree-level) approximation in perturbation theory. 
 
 \paragraph{II.}
Consider a massive scalar field which sources the Schwarzschild black hole. The energy-momentum tensor reads,
 \begin{eqnarray}
 T^{\mu\nu}_{\rm NC}(x)=\frac{1}{2}\left(\partial^{\mu}\phi \star \partial^{\nu}\phi+\partial^{\nu}\phi \star \partial^{\mu}\phi \right )-\frac{1}{2}\eta^{\mu\nu}
 \left(\partial_{\alpha}\phi \star \partial^{\alpha}\phi - m^2\phi \star \phi  \right ) \nonumber \\
 \approx T^{\mu\nu}_{0}(x) + 
 \eta^{\mu\nu}\frac{m^2\ell^4}{16}
 \theta^{\alpha\beta}\theta^{\sigma\rho}\partial_{\alpha}\partial_{\sigma}\phi\partial_{\beta}\partial_{\rho}\phi +...~,
 \label{5}
 \end{eqnarray}
  $T^{\mu\nu}_{0}=\partial^{\mu}\phi \partial^{\nu}\phi -\frac{1}{2}\eta^{\mu\nu}
 \left(( \partial \phi )^2 - m^2\phi^2  \right )$. The $\star$-product in the first line is defined as: $a(x)\star b(x)=a(x)b(x)+\sum_{n=1}^{\infty}\frac{i^n\ell^{2n}}{2^nn!}\theta^{\alpha_1\beta_1}...\theta^{\alpha_n\beta_n} \partial_{\alpha_1}...\partial_{\alpha_n}a\partial_{\beta_1}...\partial_{\beta_n}b(x)$.  In the second line of the above equation we retain only ${\cal O}(\ell^4)$ terms with four derivatives at most. This is justified in our case since we are interested in low-momentum massive particles, $m>>|\vec p|$. 
The scalar field is quantized in a standard way:
\begin{eqnarray}
\phi(x)=\int\frac{d^3\vec p}{(2\pi)^{3/2}\sqrt{2E(\vec p)}}
\left[a(\vec p){\rm e}^{-ikx}+a^+(\vec p){\rm e}^{ikx}\right ]~,  \\ \nonumber
[a(\vec p), a^+(\vec p^{\prime})]=\delta^3(\vec p-\vec p^{\prime})~,~ ~a(\vec p)|0\rangle =0~,~~ 
|\vec p\rangle = a^+(\vec p)|0\rangle ~. 
\label{6}
\end{eqnarray} 
Using (\ref{5}) and (\ref{6}), it is easy to calculate the following matrix element:
\begin{eqnarray}
\langle \vec p_2|:T^{\mu\nu}_{2}(0):|\vec p_1\rangle \approx \eta^{\mu\nu}\frac{m^2\ell^4}{16}
 \theta^{\alpha\beta}\theta^{\sigma\rho}\langle p_2|:\partial_{\alpha}\partial_{\sigma}\phi\partial_{\beta}\partial_{\rho}\phi: |p_1\rangle \nonumber \\
 = \eta^{\mu\nu}\frac{m^2\ell^4}{64\sqrt{E(\vec p_1)E(\vec p_2)}}\left(\theta^{\alpha \beta}P_{\alpha}q_{\beta}\right )^2\approx \eta^{\mu\nu}\frac{m^3\ell^4}{16}\theta^{0i}\theta^{0j}q_iq_j~,
\label{7}
\end{eqnarray}
 where in the last step we approximate $P_{\alpha}=(p_1+p_2)_{\alpha}\approx 2m\delta_{0\alpha},~E(\vec p_1), E(\vec p_2)\approx m$. Using (\ref{7}) we obtain  noncommuative corrections from (\ref{4}):
 \begin{eqnarray}
 \delta h_{00}^{\rm NC}=\frac{G_Nm^3\ell^4\theta^{0i}\theta^{0j}}{8\pi^2}\int d^3\vec q \frac{q_iq_j}{\vec q^2}{\rm e}^{i\vec q\cdot \vec r}=\frac{G_Nm^3\ell^4\theta^{0i}\theta^{0j}}{4\pi}\left[-\frac{\delta_{ij}}{r^3}+\frac{3r_ir_j}{r^5}\right ]~,  \\ \nonumber \\
 \delta h_{km}^{\rm NC}=-\delta_{km}\delta h_{00}^{\rm NC}~, \\ \nonumber \\
 \delta h_{0k}^{\rm NC}=0~.
 \label{8}
 \end{eqnarray}
 Thus the noncommutative Schwarzschild metric with the above long-distance corrections looks:
 \begin{eqnarray}
 g_{00}=\left[1-2\frac{G_Nm}{r}+2\frac{G_N^2m^2}{r^2}-2\frac{G_N^3m^3}{r^3}+...\right]+\delta h_{00}^{\rm NC}~, \\ \nonumber \\
 g_{ij}=\left[-\delta_{ij}\left( 1+2\frac{G_Nm}{r}\right)-\frac{G_N^2m^2}{r^2}\left(\delta_{ij}+\frac{r_ir_j}{r^2} \right)+2\frac{G_N^3m^3}{r^3}\frac{r_ir_j}{r^2}+... \right ]+\delta h_{ij}^{\rm NC}~,\\ \nonumber \\
 g_{0i}=0~.
  \label{9}
 \end{eqnarray}
 Recall, that the metric is given in the harmonic gauge. 
 
The first term in square brackets in (8), entering in the expression (11) for $g_{00}$,  looks pretty much the same as post-post-Newtonian correction [for $g_{ij}$ in (12) they represent third order post-Newtonian corrections] in commutative General Relativity, except the fact that the strength of this noncommutative correction is determined by $(G_N\ell^4)$ rather than by $G_N^3$.   Therefore, if $1/\ell <<M_P$, than the strength of noncommutative correction  dominate over the post-post-Newtonian one of the classical theory  by a factor $(\ell M_P)^2$. Despite this potential enhancement, noncommutative corrections are still small to be detectable in present-day experiments. Indeed, for solar system objects, a noncommutative correction is compatible with a classical post-Newtonian correction only if $\ell > 10^{-8}$ cm (atomic size). Such a large noncommutative scale is certainly excluded by particle physics observations. Interestingly, due to the second term in (8) also contributing to (11), $g_{00}$ is not a scalar anymore. This is of course attributed to the violation of  Lorentz invariance in canonical noncommutative space-times. It would be interesting to further investigate  observational effects these corrections might cause.

 \paragraph{III.} In this paper we have calculated leading semiclassical corrections to  the classical Schwarzschild metric. Our results are independent of particular realization of gravitational theory on noncommutative spacetime, since we have considered only the corrections steaming from interactions of linearized gravitational field with noncommutative matter energy-momentum tensor. 
 
These corrections are second order in noncommutative parameter, i.e. ${\cal O}(\ell^4)$ and resemble quantum corrections to classical black hole geometries calculated in commutative case in \cite{Donoghue:2001qc}. Noncommutative corrections can be significantly  larger than the quantum corrections if the scale of noncommutativity is larger than the Planck scale, $\ell >1/M_P$. Nevertheless, they are still too small to be observable in present-day experiments.
 
Unfortunately, the formalism we have used in our calculations is applicable  only for large distances from the origin of a black hole ($r>>G_Nm$). Thus our results are not capable to probe whether the geometry at the origin is singular or it is smeared as it is expected naively from the space-time noncommutativity. 
On the other hand, it would be interesting to go beyond the semiclassical approximation and calculate radiative corrections as well. If the phenomenon of UV-IR mixing, which is typical for noncommutative quantum field theories, persists also in this case, then one would expect interesting connection between large-distance corrections and small-distance geometry. Finally, the results presented here can be straightforwardly extended by considering other noncommutative fields, e.g. spinor and gauge fields, to obtain noncommutative corrections to Kerr and Reissner-Nordtrom black holes.   
 
\subparagraph{Acknowledgments.} 

I am indebted to Masud Chaichian and Anca Tureanu for email correspondence.   This  work was supported by the Australian Research Council.

\end{document}